\documentclass{rspublic}
\usepackage{graphicx}

\newcommand{\ket}[1]{\ensuremath{|#1\rangle}}
\newcommand{\bra}[1]{\ensuremath{\langle#1|}}

\newcommand{\half}{\mbox{$\frac{1}{2}$}}

\begin{document}

\title[Robust quantum logic gates]{Robust quantum information processing
with techniques from liquid state NMR}

\author[J. A. Jones]{Jonathan A. Jones}

\affiliation{Centre for Quantum Computation, Clarendon Laboratory,
University of Oxford, Parks Road, Oxford OX1 3PU, UK}

\label{firstpage}

\maketitle

\begin{abstract}{quantum logic gate, Nuclear Magnetic Resonance, composite rotation, Ising coupling}
While Nuclear Magnetic Resonance (NMR) techniques are unlikely to lead to a
large scale quantum computer they are well suited to investigating basic
phenomena and developing new techniques. Indeed it is likely that many
existing NMR techniques will find uses in quantum information processing. Here
I describe how the composite rotation (composite pulse) method can be used to
develop quantum logic gates which are robust against systematic errors.
\end{abstract}

\section{Introduction}
Nuclear Magnetic Resonance (NMR) is arguably both the best technique and the
worst technique currently known for implementing quantum information
processing.  The great strengths and weaknesses of NMR arise from the same
fundamental cause: the low frequency of NMR transitions (typically around
$500\3\textrm{MHz}$, corresponding to about $2\3\mu\textrm{eV}$). This makes
NMR experiments easy to perform (the experimental timescale is conveniently
slow), and also acts to minimise the effects of decoherence. However the
extreme weakness of NMR signals means that it is not yet possible to study
single nuclear spins: instead we must use macroscopic ensembles, which occupy
hot thermal states.

These strengths and weaknesses mean that, while it is extremely unlikely that
liquid state NMR techniques will ever be used to construct a large scale
general purpose quantum computer (Jones 2000), NMR provides an excellent
technique for conducting preliminary studies (Cory \textit{et al.} 1996, 1997;
Gershenfeld \& Chuang 1997; Jones \& Mosca 1998; Chuang \textit{et al.} 1998),
and for developing techniques which will be used in large scale devices.
Furthermore, the NMR community has developed a sophisticated library of
techniques for manipulating nuclear spins (Ernst \textit{et al.} 1987; Freeman
1997; Claridge 1999), many of which can be directly transferred to manipulate
qubits in other implementations.

In this paper I discuss the method of composite rotations, also called
composite pulses (Levitt 1986), which are widely used in NMR to combat
systematic errors arising from inevitable experimental imperfections.  While
many composite pulses developed for use in NMR cannot be directly transferred
to quantum computing some can be, and novel composite pulses have been
developed specifically for use in quantum computing (Cummins \& Jones 2000;
Cummins \textit{et al.} 2002). More recently the concept of composite
rotations has been extended to two-qubit (controlled) logic gates (Jones
2002). Although these robust quantum logic gates have been developed in the
context of NMR, and are described here using NMR terminology, the basic ideas
are entirely general and can be used in many other experimental
implementations.

\section{Spins, qubits and the Bloch sphere}
The majority of conventional NMR studies are conducted on nuclei with spin
$I=1/2$, and these also provide a natural method of implementing quantum
information processing, as the two spin states, \ket{\alpha} and \ket{\beta},
can be trivially mapped to the two basis states of a qubit, \ket{0} and
\ket{1}.  A general superposition state (Nielsen \& Chuang 2000) can be
written as
\begin{equation}
\ket{\psi}=\cos(\theta/2)\ket{0}+\sin(\theta/2)\re^{i\phi}\ket{1}
\end{equation}
(neglecting irrelevant global phases) and so can be thought of as a point on a
unit sphere, traditionally called the Bloch sphere, with states \ket{0} and
\ket{1} at the north and south poles, and the equally weighted superpositions
lying around the equator.  Any unitary operation on a single isolated qubit
(any single qubit logic gate) corresponds to a rotation of this sphere.

Single qubit rotations can be specified by their rotation axis and their
rotation angle; the rotataion axis can itself be described by a single point
on the sphere.  Within the NMR literature spin states and unitary operations
are both described using the product operator notation (S{\o}rensen \textit{et
al.} 1983; Ernst \textit{et al.} 1987; Hore \textit{et al.} 2000) . For a
single spin
\begin{equation}
\begin{split}
\ket{\psi}\bra{\psi}&=\frac{1}{2}\left(\textbf{1}+
\begin{pmatrix}
\cos\theta & \sin\theta\,\re^{-i\phi} \\
 \sin\theta\,\re^{i\phi} & -\cos\theta
\end{pmatrix}
\right)\\
&=\frac{1}{2}\,(\sigma_{0}+\sin\theta\,\cos\phi\,\sigma_{x}+\sin\theta\,\sin\phi\,\sigma_{y}+\cos\theta\,\sigma_{z})\\
&=\half{E}+\sin\theta\,\cos\phi\,I_x+\sin\theta\,\sin\phi\,I_y+\cos\theta\,I_z
\end{split}
\end{equation}
where the product operators used in the final line are closely related to the
corresponding Pauli matrices.  NMR systems are usually in hot thermal
ensembles, and so are not described by pure states but by highly mixed states
(Jones 2001); for a single spin the thermal equilibrium state take the simple
form
\begin{equation}
\rho=\textbf{1}+\epsilon I_z
\end{equation}
It is customary to neglect the \textbf{1} term (which cannot be observed by
NMR techniques) and the $\epsilon$ factor (which simply determines the signal
intensity) and so the equilibrium state is described as $I_z$.

\section{Pulses and pulse errors}
NMR experiments are composed of a series of radiofrequency (RF) pulses, which
cause rotations about axes in the $xy$-plane, and periods of free evolution,
which for a single spin can be described as $z$ rotations.  Two particularly
common pulses (Jones 2001) are the $90^\circ$ pulse corresponding to
excitation
\begin{equation}
I_z\xrightarrow{\re^{-i\pi/2\,I_y}}I_x \label{eq:p90y}
\end{equation}
(closely related to the Hadamard gate) and the $180^\circ$ pulse corresponding
to inversion
\begin{equation}
I_z\xrightarrow{\re^{-i\pi I_x}}-I_z \label{eq:p180x}
\end{equation}
(in effect, a NOT gate).  In each case the rotation angle is determined by the
applied RF power (which determines the rotation rate) and the length of time
over which the RF is applied, while the phase of the rotation axis (in the
$xy$ plane) is determined by the phase of the RF field.

Genuine experimental implementations are, of course, not quite as perfect as
the description above implies.  Clearly a rotation can go wrong in one of two
ways: there can be errors in the rotation angle or in the rotation axis.  The
first type of error, usually called a pulse length error, typically arises
when the RF field strength deviates from its assumed value, so that all
rotations are systematically too long or too short by some constant fraction.
Errors of the second kind, usually called off-resonance effects, occur when
the RF field is not exactly in resonance with the transition, resulting in
evolution around an effective field, tilted out of the $xy$ plane towards the
$\pm z$ axis.  The unitary operation describing a real pulse therefore takes
the form
\begin{equation}
U=\exp\left(-i\times2\pi\,\nu\times[(1+g)(I_x\,\cos\phi+I_y\,\sin\phi)+fI_z]\times{t}\right)
\end{equation}
where $\nu$ is the nominal rotation rate, $t$ is the pulse length, $\phi$ is
the pulse phase, $g$ is the fractional error in the RF power, and $f$ is the
off-resonance fraction, given by $f=\delta/\nu$, where $\delta$ is the
off-resonance frequency error.

The ideal inversion pulse, equation~(\ref{eq:p180x}), occurs when
$\nu{t}=\pi$, $\phi=0$, and $f=g=0$; in real pulses the last two conditions
are relaxed.  While both errors can (and do) occur simultaneously, one is
typically dominant, and it is most useful to begin by considering these errors
separately.  Furthermore conventional NMR experiments begin in some well
defined initial state, usually $I_z$, and the quality of a pulse can be
assessed by the overlap between the final state and its ideal form (for an
inversion pulse, $-I_z$): it is not necessary (or even desirable) to consider
the effects of the pulse on other initial states.

\begin{figure}
\begin{center}
\includegraphics[scale=0.5]{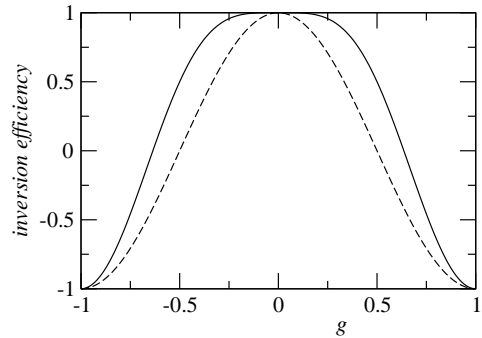}
\end{center}
\caption{Performance of naive (dashed line) and $90^\circ_y 180^\circ_x
90^\circ_y$ composite (solid line) inversion pulses in the presence of pulse
length errors. Performance is assessed by the component of the final state
lying along the $-I_z$ axis.} \label{fig:invertg}
\end{figure}

In the presence of pulse-length errors, an inversion pulse performs the
transformation
\begin{equation}
I_z\xrightarrow{\re^{-i\pi(1+g)I_x}}-\cos(\pi g)I_z+\sin(\pi g)I_y
\label{eq:p180xg}
\end{equation}
and the component of the final state along the $-I_z$ axis, $\cos(\pi g)$,
provides a convenient quality measure, lying in the range $\pm 1$.  This
function is plotted in figure~\ref{fig:invertg}; clearly the sequence only
performs well for very small values of $g$ (near perfect pulses).
The conventional NMR method for dealing with pulse length errors in inversion
pulses is to replace the simple $180^\circ_x$ pulse with the composite pulse
sequence $90^\circ_y 180^\circ_x 90^\circ_y$.  This sequence has an inversion
efficiency given by
\begin{equation}
\cos(\pi g)+\half\sin^2(\pi g)
\end{equation}
which is also plotted in figure~\ref{fig:invertg}.  The composite sequence
performs better than the naive sequence for essentially all values of $g$, but
especially for moderate errors, in the range $\pm10\%$.  The manner in which
this improvement is achieved can be understood by examining the trajectory on
the Bloch sphere in the presence of small errors (Freeman 1997; Claridge
1999).

The situation in the presence of off-resonance effects is more complex, as
both the rotation axis and rotation angle are affected.  Composite pulse
sequences are known which tackle these effects (Levitt 1986), but this point
will not be explored further here.

\section{Pulses and logic gates}
Inversion, which takes $I_z$ to $-I_z$ is clearly closely related to the NOT
gate, which takes \ket{0} to \ket{1}, but the two processes are not simply
equivalent (Jones 2001).  The NOT gate corresponds to a $180^\circ$ rotation
around the $I_x$ axis, while inversion can be achieved by a $180^\circ$
rotation around \textit{any} axis in the $xy$ plane.  The difference is that
an inversion sequence need \textit{only} act correctly on the initial states
\ket{0} and \ket{1} (corresponding to $I_z$ and $-I_z$), but a NOT gate must
also act correctly on any superposition of these states.  In NMR terms this
means that the gate must also interchange $I_y$ and $-I_y$, and must leave
$\pm I_x$ unchanged. It is, therefore, important to analyse the composite
inversion sequence, $90^\circ_y 180^\circ_x 90^\circ_y$, to see how it
performs with these initial states.

In the absence of pulse length errors the composite pulse sequence does, in
fact, perform correctly, but in the presence of errors the situation is not so
good.  While the composite pulse sequence performs better than a simple
$180^\circ_x$ pulse for $\pm I_z$ states, it performs \textit{worse} than the
simple sequence for $\pm I_x$ states; the effects of the two sequences on $\pm
I_y$ states are identical.  This behaviour is exactly what one might expect:
it seems intuitively reasonable that composite pulse sequences should
redistribute errors over the Bloch sphere, rather than actually reduce them
(Cummins 2001). If this were indeed the case, then composite pulses would have
little to offer quantum information processing, but surprisingly some
composite pulses are known which perform well for \textit{all} initial states.
Such sequences, sometimes called Class A composite pulses (Levitt 1986), are
of little use in conventional NMR, and so have received relatively little
study.  They are, however, ideally suited to implementing quantum logic gates.

The first application of composite pulses to quantum information processing
was by Cummins \& Jones (2000), who used composite $90^\circ$ pulses to reduce
the influence of off-resonance effects on an implementation of quantum
counting. More recently Cummins \textit{et al.} (2002) have described two
families of composite pulse sequences which correct for pulse length errors.
From here I shall concentrate on one of these, the BB1 sequence originally
developed by Wimperis (1994).

\section{The BB1 composite pulse sequence}
The BB1 composite pulse sequence was developed with two principal aims:
firstly to provide good compensation for pulse length errors, and secondly to
provide a composite pulse sequence which could be used to replace any simple
pulse at any position in a pulse sequence (Wimperis 1994).  The second aim is
essentially equivalent to seeking a Class A composite pulse, and BB1 does
indeed have this property.  The first aim is also well achieved by BB1, which
provides a quite remarkable degree of compensation for pulse length errors: it
is not only better than any other known Class A composite pulse, it can also
provide better compensation that many conventional sequences tailored to
specific operations (such as inversion).

When assessing the quality of a Class A composite pulse it is necessary to
determine how well the unitary transformation actually implemented ($V$)
approximates the desired unitary transformation ($U$).  A simple and
convenient definition of this fidelity is given by
\begin{equation}
\mathcal{F}=\frac{|\text{Tr}(VU^\dag)|}{\text{Tr}(UU^\dag)} \label{eq:fidp}
\end{equation}
(note that it is necessary to take the absolute value of the numerator as $U$
and $V$ could in principle differ by an irrelevant global phase shift).  A
simpler approach, appropriate to single qubit logic gates, is to note that any
unitary operation on a single qubit is a rotation, and so can be represented
by a quaternion
\begin{equation}
\mathsf{q}=\{s, \mathbf{v}\}
\end{equation}
where
\begin{equation}
s=\cos(\theta/2)
\end{equation}
depends solely on the rotation angle, $\theta$, and
\begin{equation}
\mathbf{v}=\sin(\theta/2)\mathbf{a}
\end{equation}
depends on both the rotation angle, $\theta$, and a unit vector along the
rotation axis, $\mathbf{a}$.  The quaternion describing a composite pulse
sequence is obtained by multiplying the quaternions for each pulse according
to the rule
\begin{equation}
\mathsf{q_1}\ast\mathsf{q_2}=\{s_1\cdot s_2 -\mathbf{v_1}\cdot\mathbf{v_2},
s_1\mathbf{v_2}+s_2\mathbf{v_1}+\mathbf{v_1}\wedge\mathbf{v_2}\}
\end{equation}
while two quaternions can be compared using the quaternion fidelity (Levitt
1986)
\begin{equation}
\mathcal{F}(\mathsf{q_1},\mathsf{q_2})=|\mathsf{q_1}\cdot\mathsf{q_2}|
=|{s_1}\cdot{s_2}+\mathbf{v_1}\cdot\mathbf{v_2}| \label{eq:fidq}
\end{equation}
(it is necessary to take the absolute value, as the two quaternions
$\{s,\mathbf{v}\}$ and $\{-s,-\mathbf{v}\}$ correspond to equivalent
rotations, differing in their rotation angle by integer multiples of $2\pi$).
For single qubit operations the two fidelity definitions (equations
\ref{eq:fidp} and \ref{eq:fidq}) are equivalent, and quaternions will be used
from here on.

I shall take as my target operation a NOT gate, that is a $180^\circ_x$
rotation; similar results can be obtained for any other desired rotation
(Wimperis 1994; Cummins \textit{et al.} 2002).  Thus the quaternion
representing the ideal operation is
\begin{equation}
\mathsf{q_0}=\{0,(1,0,0)\}
\end{equation}
while the quaternion representing the rotation which actually occurs (as a
result of pulse length errors) is
\begin{equation}
\mathsf{q_1}=\{\cos[(1+g)\pi/2],(\sin[(1+g)\pi/2],0,0)\}
\end{equation}
giving rise to a quaternion fidelity of
\begin{equation}
\mathcal{F}_1=\cos(g\pi/2)\approx 1-\frac{\pi^2g^2}{8}
\end{equation}
(this expression neglects taking the absolute value, and so is only valid for
values of $g$ in the range $\pm100\%$).  The conventional composite pulse
sequence $90^\circ_y 180^\circ_x 90^\circ_y$ which has the quaternion form
\begin{equation}
\mathsf{q_2}=\{\sin^2[g\pi/2],(\cos[g\pi/2],-\sin[g\pi]/2,0)\}
\end{equation}
gives exactly the same fidelity, $\mathcal{F}_2=\cos(g\pi/2)=\mathcal{F}_1$.
This confirms that the conventional sequence does not actually correct for
errors, but simply redistributes them (Cummins 2001).

One BB1 version of a NOT gate takes the form $90^\circ_0 180^\circ_{\phi_1}
360^\circ_{\phi2} 180^\circ_{\phi_1} 90^\circ_0$, where the phase angles
$\phi_1$ and $\phi_2$ remain to be determined, and a phase angle of $0$
corresponds with an $x$ rotation.  Note that this composite pulse sequence
comprises a cluster of $180^\circ$ and $360^\circ$ pulses placed in the middle
of a $180^\circ_0$ pulse.  In the absence of errors the central cluster has no
effect whatsoever, and the pulse sequence collapses to a simple $180^\circ_0$
pulse.  In the presence of pulse length errors the central cluster will have
some effect, and the intention is to choose values of $\phi_1$ and $\phi_2$
such that the effects of the central cluster compensate the errors in the
outer pulses.  Note that the sequence discussed here differs subtly from the
original BB1 sequence described by Wimperis (1994) which had the cluster
placed \textit{before} the $180^\circ_0$ pulse; in fact it can be shown that
the cluster may be placed at any point with respect to this pulse without
affecting the fidelity (Cummins \textit{et al.} 2002).

\begin{figure}
\begin{center}
\includegraphics[scale=0.5]{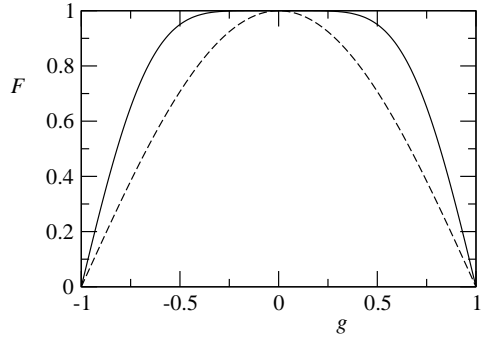}
\end{center}
\caption{Quaternion fidelity of naive (dashed line) and BB1 composite (solid
line) $180^\circ_x$ pulses used to implement NOT gates in the presence of
pulse length errors. The BB1 sequence outperforms a simple pulse for all pulse
length errors in the range $\pm100\%$, and is particularly good in the range
$\pm25\%$.} \label{fig:BB1}
\end{figure}

The quaternion describing the BB1 composite pulse sequence is complicated. Its
$z$ component is zero, as expected for a time-symmetric pulse sequence
(Cummins \textit{et al.} 2002), but the remaining components show a complex
dependence on $\phi_1$, $\phi_2$ and $g$.  Progress is most easily made by
expanding the quaternion as a Maclaurin series in $g$.  The first order $y$
component can be set to 0 by choosing $\phi_2=3\phi_1$, leaving the
approximate quaternion
\begin{equation}
\mathsf{q_3}\approx\{-\half\pi[1+4\cos\phi_1]g,(1,0,0)\}
\end{equation}
(neglecting terms $\textrm{O}(g^2)$ and higher). Finally the scalar part of
$\mathsf{q_3}$ can be made approximately equal to 0 by choosing
$\phi_1=\pm\arccos(-1/4)$, and following previous practice the positive
solution is taken.  These choices result in a quaternion with a fidelity
\begin{equation}
\mathcal{F}_3=\frac{1}{128}\left(150\cos(g\pi/2)-25\cos(3g\pi/2)+3\cos(5g\pi/2)\right)\approx1-\frac{5\pi^6g^6}{1024}
\end{equation}
in which \textit{both} the second and fourth order error terms have been
cancelled. It is clear from this analysis that the BB1 composite pulse
outperforms a simple $180^\circ$ pulse for small values of $g$.  In fact it
does better for \textit{all} values of $g$ in the range $\pm100\%$, as shown
by the fidelity plots in figure~\ref{fig:BB1}; the most spectacular effects
are seen within the range $\pm10\%$, as shown in table~\ref{table:BB1}.  BB1
pulses can implement extremely accurate gates in the presence of moderate
errors (1--10\%), while naive pulses require impossibly accurate control of
the RF field strength to achieve the same quality.
\begin{table}
\caption{Quaternion infidelities ($1-\mathcal{F}$) of naive and BB1 composite
$180^\circ_x$ pulses.}\label{table:BB1}
\begin{center}
\begin{tabular}{lll}
\hline $g$ & naive & BB1 \\\hline
0.1&$1.2\times10^{-2}$&$4.6\times10^{-6}$\\
0.03&$1.1\times10^{-3}$&$3.4\times10^{-9}$\\
0.01&$1.2\times10^{-4}$&$4.7\times10^{-12}$\\
0.003&$1.1\times10^{-5}$&$3.4\times10^{-15}$\\
0.001&$1.2\times10^{-6}$&$4.7\times10^{-18}$\\
\hline
\end{tabular}
\end{center}
\end{table}

It is also instructive to examine the effect of the BB1 pulse on particular
initial states.  For initial states lying along \textit{any} of the cardinal
axes the BB1 sequence results in an error term of order $g^6$, although the
exact size of the term depends on the choice of axis.  In comparison, for
initial states along $\pm I_z$ a simple $180^\circ_x$ pulse results in an
error of order $g^2$, while the conventional composite pulse sequence,
$90^\circ_y 180^\circ_x 90^\circ_y$, gives an error of order $g^4$; thus the
BB1 sequence acts as a better inversion sequence than the conventional
composite pulse sequence designed to perform an inversion! For initial states
along $\pm I_y$ the conventional composite pulse sequence provides no
compensation, and both it and and the simple pulse give errors of order $g^2$.
The only blemish on the BB1 sequence is seen when examining initial states
along $\pm I_x$, for which the simple pulse performs perfectly (the
conventional composite pulse gives an error of order $g^2$).  This property of
perfect behaviour along one single axis is a particular property of simple
pulses, and cannot be achieved with composite pulses.  The very best behaviour
for BB1 is observed for initial states along two particular axes in the $xz$
plane, for which an error of order $g^{10}$ is seen.

While the performance of BB1 is extremely impressive, it would obviously be
desirable to find an even better sequence, with even better error tolerance.
Although such sequences probably exist it is not clear how they can be found.
Initial attempts in this direction (G. Llewellyn, unpublished results) have
had no success, but have simply made clear how unusually good BB1 actually is.

Very similar composite pulses can be obtained for other pulse angles (Wimperis
1994; Cummins \textit{et al.} 2002): a $\theta_0$ pulse is replaced by
$(\theta/2)_0 180^\circ_{\phi}360^\circ_{3\phi}180^\circ_{\phi} (\theta/2)_0$
with $\phi=\arccos(-\theta/4\pi)$.  There is, however, a subtle point
concerning the accuracy with which such pulse sequences may be implemented.
Typically all the pulses in such a sequence are implemented by applying the
same RF field for different lengths of time, and the clock controlling the RF
field has a finite time resolution.  While it is not necessary to control the
absolute lengths of each pulse to very great accuracy, it is
\textit{essential} that the relative lengths of each pulse are correct.  This
is easily achieved when $\theta$ is $180^\circ$ or some simple fraction of it,
as all the pulses can then be implemented as multiples of some common element,
but is much more difficult for arbitrary angles.

\section{Two qubit logic gates}
It is well known that any desired circuit can be constructed using single
qubit logic gates in combination with any one non-trivial two qubit logic gate
(Deutsch \textit{et al.} 1995).  The two qubit gate most commonly discussed is
the controlled-NOT gate (Barenco \textit{et al.} 1995) which applies a
$180^\circ_x$ rotation to its target qubit conditional on its control qubit
being in the state \ket{1}. An essentially equivalent, and frequently more
convenient, alternative is the controlled-phase gate, which applies a
$180^\circ_z$ rotation to its target qubit conditional on the state of its
control qubit. Note that in this case the logic gate acts symmetrically on the
two qubits: the control/target distinction is convenient but artificial.  This
choice of two qubit gate is particularly convenient in implementations, such
as NMR, built around Ising couplings, as the controlled-phase gate and
evolution under the Ising coupling are trivially related (Jones 2001, 2002). A
controlled-NOT gate can then be implemented by applying Hadamard gates to the
target nucleus before and after the controlled-phase gate.

Consider a system of two spin-$1/2$ nuclei, $I$ and $S$.  The Ising coupling
gate is implemented by evolution under the $J$ coupling Hamiltonian
\begin{equation}
\mathcal{H}_{IS}=\pi J\, 2I_zS_z
\end{equation}
for a time $\tau=\phi/\pi J$, where $J$ is the coupling strength and $\phi$ is
the desired evolution angle.  The desired controlled-phase gate requires
$\phi=\pi/2$ and so $\tau=1/2J$; in NMR this is known as the antiphase
condition. In order to implement accurate controlled-phase gates it is clearly
necessary to know $J$ with corresponding accuracy.  This is relatively simple
in NMR studies of small molecules, but is much more difficult in larger
systems. In particular, many experimental proposals contain an array of qubits
coupled by Ising interactions (Ioffe \emph{et al.} 1999; Cirac \& Zoller 2000;
Briegel \& Raussendorf 2001; Raussendorf \& Briegel 2001), with $J$ couplings
that are nominally identical but in fact differ from one another as a result
of imperfections in the lattice. In systems of this kind it is desirable to be
able to perform some accurately known Ising evolution over a \textit{range} of
values of $J$. Perhaps surprisingly, this is relatively easy to achieve using
composite pulse techniques.

\begin{figure}
\begin{center}
\includegraphics{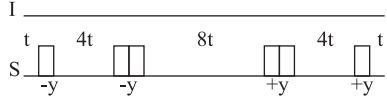}
\end{center}
\caption{Pulse sequence for a BB1 based robust Ising gate to implement a
controlled-phase gate. Boxes correspond to single qubit rotations with
rotation angles of $\phi=\arccos(-1/8)\approx97.2^\circ$ applied along the
$\pm y$ axes as indicated; time periods correspond to evolution under the
Ising coupling, $\pi J\,2I_zS_z$, for multiples of the time $t=1/4J$.  The
simple Ising gate corresponds to free evolution for a time $2t$.}
\label{fig:IsingPulses}
\end{figure}

The problem of performing accurate Ising evolutions is conceptually similar to
that of correcting for pulse length errors in single qubit gates, and the
solutions are closely related (Jones 2002).  Ising coupling corresponds to
rotation about the $2I_zS_z$ axis, and errors in $J$ correspond to errors in
the rotation angle about this axis.  These can be parameterised by the
fractional error in the value of $J$:
\begin{equation}
g=\frac{J_{\textrm{real}}}{J_{\textrm{nominal}}}-1.
\end{equation}
Errors of this kind can be overcome by rotating about a sequence of axes
tilted from $2I_zS_z$ towards another axis, such as $2I_zS_x$. Defining
\begin{equation}
\theta_\phi\equiv\exp[-i\times\theta\times(2I_zS_z\cos\phi+2I_zS_x\sin\phi)]
\end{equation}
allows the direct evolution sequence $(\pi/2)_0$ to be replaced by the
composite pulse sequence $(\pi/4)_0 (\pi)_{\phi} (2\pi)_{3\phi} (\pi)_{\phi}
(\pi/4)_0$ with $\phi=\arccos(-1/8)$.  The tilted evolutions can be realised
(Ernst \textit{et al.} 1987) by sandwiching a $2I_zS_z$ rotation (free
evolution under the Ising Hamiltonian) between $\phi_{\mp y}$ pulses applied
to spin $S$. After cancellation of extraneous pulses the final sequence takes
the form shown in figure~\ref{fig:IsingPulses}.
Note that the labelling of the two spins as $I$ and $S$ is arbitrary, and the
$\phi_{\mp y}$ pulses can be applied to the other spin if this is more
convenient.

It is vital that any robust implementation of a quantum logic gate be built
from components that are themselves robust.  The robust Ising gate uses only
two components: single qubit rotations around the $\pm y$ axes, for which
robust versions are described above, and periods of evolution under the Ising
coupling.  As before it is not necessary to accurately control the absolute
lengths of the five time periods, but they must have lengths in the integer
ratios $1:4:8:4:1$.

\begin{figure}
\begin{center}
\includegraphics[scale=0.5]{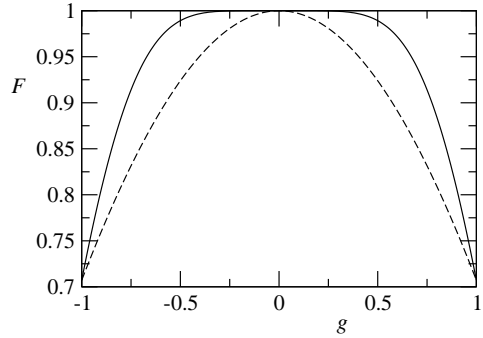}
\end{center}
\caption{Propagator fidelity of simple (dashed line) and BB1 robust (solid
line) Ising gates used to implement controlled-phase gates. The BB1 sequence
outperforms a simple gate for errors in the range $\pm100\%$; over the range
$\pm25\%$ the robust gate is indistinguishable from perfection on this scale.}
\label{fig:BB1J}
\end{figure}

The fidelity gain achieved for Ising coupling gates by this approach is
\textit{identical} to that achieved for single qubit rotations.  (For two
qubit gates it is necessary to use the propagator fidelity,
equation~\ref{eq:fidp}, but as mentioned above this is equivalent to the
quaternion fidelity in the single qubit case).  As the controlled-phase gate
corresponds to a $90^\circ$ rotation, rather than the $180^\circ$ rotations
discussed in the case of single qubit gates, the fidelities are slightly
different from those discussed previously.  For a simple Ising gate
\begin{equation}
\mathcal{F}\approx 1-\frac{\pi^2g^2}{32}
\end{equation}
while for a BB1 based robust Ising gate
\begin{equation}
\mathcal{F}\approx 1-\frac{63\pi^6g^6}{65536}.
\end{equation}
The fidelities are plotted in figure~\ref{fig:BB1J}.
Clearly the robust Ising gate compensates well for small errors in $J$ values,
especially within the range $\pm25\%$.  Over the range $\pm10\%$ the
infidelity of the robust gate is always less that one part in $10^6$; to
achieve comparable fidelity with a simple gate it is necessary to control $J$
to better than $0.2\%$, more than 50 times more accurately then is needed for
the robust gate.

\section{Conclusions}
Composite rotations show great promise as a method for combatting systematic
errors in quantum logic gates.  Without progress in this area attempts to
build large scale quantum computing devices will founder on the need for
impossibly precise experimental control.  Methods have been derived for
tackling both pulse length errors and off-resonance effects in single qubit
gates (Cummins \& Jones 2000; Cummins \textit{et al.} 2002) and for tackling
variations in the coupling strength in the Ising coupling two-qubit gate
(Jones 2002). Together these provide a universal set of robust quantum logic
gates within the Ising coupling model. Although developed within the context
of NMR, Ising couplings play a major role in many proposed implementations of
quantum information processing (Ioffe \emph{et al.} 1999; Cirac \& Zoller
2000; Briegel \& Raussendorf 2001; Raussendorf \& Briegel 2001), and these
robust gates are likely to find their final applications elsewhere.

\begin{acknowledgements}
I thank the Royal Society of London for financial support.  I am grateful to
S.~Benjamin, H.~K. Cummins, L.~Hardy, G.~Llewellyn, M.~Mosca and A.~M. Steane
for helpful discussions.
\end{acknowledgements}

\label{lastpage}
\end{document}